\documentclass[12pt]{article}
\usepackage{hyperref}
\usepackage[fleqn]{amsmath}
\usepackage{amssymb} % e.g. \gtrsim
\usepackage{bbm} % BlackBoeard letters
\usepackage[small]{caption2} % font of captions is different from text
\usepackage{fleqn} % equations are not centered
\usepackage{graphicx} % \including PostScript
\usepackage{mathrsfs}   % e.g. nice Lagrange-L with \mathscr{L}
\usepackage[small,loose]{subfigure}  % subfigures with (a), (b) etc., ... and own caption
\usepackage{cite} % [1-5] instead of [1,2,3,4,5]
\usepackage{color}
\usepackage{cancel}

\newcommand{\eVdist}{\kern-0.06em}
\newcommand{\rep}[1]{\ensuremath\boldsymbol{#1}}
\newcommand{\crep}[1]{\ensuremath\overline{\boldsymbol{#1}}}

\newcommand{\SO}[1]{\ensuremath{\mathrm{SO}(#1)}}
\newcommand{\SU}[1]{\ensuremath{\mathrm{SU}(#1)}}
\newcommand{\U}[1]{\ensuremath{\mathrm{U}(#1)}}
\newcommand{\Z}[1]{\ensuremath{\mathbbm{Z}_{#1}}} % z_N ->\Z{N}

\numberwithin{equation}{section}
\numberwithin{table}{section}

\allowdisplaybreaks[1]

\usepackage{nicefrac}
\renewcommand{\frac}{\nicefrac}
\title{No--go theorems for anomaly--free discrete symmetries in four--dimensional GUTs
}

\begin{document}

\begin{titlepage}

\vspace*{-3.0cm}
\begin{flushright}
TUM-HEP 817/11\\ CERN-PH-TH/2011-230
\end{flushright}

\vspace*{1.0cm}

\begin{center}
{\Large\bf
No--go theorems for $\boldsymbol{R}$ symmetries}\\[0.2cm]
{\Large\bf in four--dimensional GUTs
}

\vspace{1cm}

\textbf{
Maximilian~Fallbacher$^{a}$,
Michael~Ratz$^{a,b}$, Patrick~K.S.~Vaudrevange$^{a}$
}
\\[8mm]
\textit{$^a$\small
~Physik--Department T30, Technische Universit\"at M\"unchen, \\
James--Franck--Stra\ss e, 85748 Garching, Germany
}
\\[5mm]
\textit{$^b$\small
~Theory Group, CERN, 1211 Geneva 23, Switzerland
}
\end{center}

\vspace{1cm}

\begin{abstract}
We prove that it is impossible to construct a grand unified model,
based on a simple gauge group, in four dimensions that leads to the exact
MSSM, nor to a singlet extension, and possesses an unbroken $R$ symmetry. This
implies that no MSSM model with either a $\Z{M\ge3}^R$ or $\U1_R$ symmetry can
be completed by a four--dimensional GUT in the ultraviolet. However, our no--go
theorem does not apply to GUT models with extra dimensions. We also show
that it is impossible to construct a 4D GUT that leads to the MSSM plus an
additional anomaly--free symmetry that forbids the $\mu$ term.
\end{abstract}

\end{titlepage}

\newpage

%%%%%%%%%%%%%%%%%%%%%%%%%%%%%%%%%%%%%%%%%%%%%%%%%%%%%%%%%%%%%%%%%%%%%%%%%%%%%%%%%%%%%%
\section{Introduction}
%%%%%%%%%%%%%%%%%%%%%%%%%%%%%%%%%%%%%%%%%%%%%%%%%%%%%%%%%%%%%%%%%%%%%%%%%%%%%%%%%%%%%%

The scheme of supersymmetric grand unification provides an attractive framework
for physics beyond the standard model (SM) of particle physics. Apart from the
observation that gauge couplings seem to unify at a scale of a few times
$10^{16}\,\mathrm{GeV}$ \cite{Dimopoulos:1981yj} in the minimal
supersymmetric standard model (MSSM), the structure of matter hints at
unification. SM matter comes in three copies of $\rep{10}\oplus\crep{5}$
representations under
\begin{equation}
 \SU5~\supset~\SU3_\mathrm{C}\times\SU2_\mathrm{L}\times\U1_\mathrm{Y}~=~G_\mathrm{SM}\;,
\end{equation}
or, after introducing the right--handed neutrino, in form of three
$\rep{16}$--plets of \SO{10}. Arguably, the most compelling explanation of
the smallness of neutrino masses is due to the see--saw mechanism
\cite{Minkowski:1977sc}, which also appears to require the see--saw scale to be
close to $M_\mathrm{GUT}$. However, despite all these hints the scheme of
grand unified theories (GUTs) does not yet provide us with a clear picture. For
instance, typical obstacles encountered when constructing GUTs in four
dimensions include the so--called doublet--triplet splitting problem, i.e.\ the
question why Higgs fields appear in split multiplets, and, associated to it, the
prediction of too fast proton decay. 

While, arguably, all known proposals for doublet--triplet splitting in
four--dimensional (4D) GUTs have some weak points, up to now there exists no
argument for why this is necessarily the case. One purpose of this letter is to
give such an argument.

Suppose there is indeed a doublet--triplet splitting mechanism which can be
completely understood in terms of 4D physics. Then one should be able to
understand in the effective theory why the $\mu$ term essentially vanishes. If
the smallness of the $\mu$ term is to be `natural' (in 't Hooft's sense
\cite{'tHooft:1979bh}), there has to be a symmetry that forbids it. On the other
hand, it has been shown that, if one demands consistency with grand unification
and anomaly freedom, then only $R$ symmetries may forbid the $\mu$ term
\cite{Lee:2010gv} (cf.\ also the somewhat similar discussion in
\cite{Hall:2002up}). It therefore appears that, if one is to solve the $\mu$
problem in `a natural way', $R$ symmetries are instrumental.

However, we shall prove that for a spontaneously broken GUT symmetry
(based on a simple Lie group) in four dimensions one cannot
get the exact MSSM with residual $R$ symmetries. This allows us to
conclude that a `natural' solution to the doublet--triplet problem is not
available in four dimensions. Our proof applies to singlet extensions of the
MSSM as well and, in what follows, we will use the abbreviation MSSM also for
these singlet extensions. 

This Letter is organized as follows. We will start with the special case of a
$\SU5\times \Z{M}^R$ in section \ref{sec:NoGoZM} and extend the result obtained
there to more general cases in section \ref{sec:furtherNoGo}. Implications of
our no--go theorem for model building are discussed in section
\ref{sec:applications}. Section \ref{sec:ExtraDims} is devoted to the question
of circumventing our no--go theorem in extra dimensions while section
\ref{sec:summary} contains our summary.

\section{No $\boldsymbol{R}$ symmetries from 4D GUTs}
\label{sec:NoGoZM}

This section is devoted to the proof that it is impossible to construct a GUT
(based on a simple gauge group) in four dimensions with a finite
number of multiplets that leads to the MSSM (or any of its singlet extensions)
with a residual $R$ symmetry. While our discussion is based on Abelian discrete
$R$ symmetries, denoted by $\Z{M}^R$ in what follows, it also applies to
continuous, i.e.\ $\U1_R$, symmetries because in this case one can always resort
to a discrete subgroup $\Z{M}^R \subset \U1_R$. In our analysis, we focus on
discrete $\Z{M}^R$ symmetries with $M\geq3$.\footnote{Discrete $R$ symmetries of
order two are no `true' $R$ symmetries since any global supersymmetric theory
possesses a symmetry under which the superspace coordinates transform as
$\theta\to-\theta$ and all spin--$\nicefrac{1}{2}$ fermions get multiplied by
$-1$. In particular, using this `automatic' symmetry one can easily convince
oneself that the so--called $R$ parity of the MSSM \cite{Fayet:1977yc} is
equivalent to matter parity \cite{Dimopoulos:1981dw} (cf.\ also the discussion
in \cite{Dine:2009swa}).}  Our conventions are such that the superpotential
carries $R$ charge 2. We start by discussing 4D \SU5 GUTs in \ref{sec:SU5}, and
then consider generalizations in \ref{sec:furtherNoGo}.

\subsection{Massless exotics vs.\ unbroken $\boldsymbol{\Z{M}^R}$ in 4D
SU(5) GUTs}
\label{sec:SU5}

Consider the MSSM with an additional $\Z{M}^R$ symmetry, i.e.\ the symmetry
group of the model is $G_\mathrm{SM}\times \Z{M}^R$ (possibly amended by further
symmetries). We can then ask whether this symmetry group can emerge from an \SU5
GUT by spontaneous breaking; the extension to larger GUT groups is deferred to
\ref{sec:LargerGUTs}. Since there is a residual $\Z{M}^R$ symmetry, the
symmetries at the GUT level have to contain $\Z{M}^R$ as a subgroup. Without
loss of generality, we can base our discussion on a GUT with $\SU5\times\Z{M}^R$
symmetry, although the actual ($R$ and/or non-$R$) symmetry before spontaneous
breaking might be larger (and lead to stronger conditions than we need). In
other words, in case there is actually a larger symmetry group above the GUT
scale, the charges we will refer to will always be the ones of the $\Z{M}^R$
subgroup.

We proceed by classifying the GUT multiplets $\rep{R}$ according to their
$\Z{M}^R$ charges. Most mass terms between such multiplets are prohibited by
$\Z{M}^R$. For our purposes it will be sufficient to focus on the subsector of
fields with charges 0 and 2. Particles of the latter two types can only have
mass terms of the form $\mathcal{M}(m,\langle H_0 \rangle,
\Lambda)~\Psi_0~\Phi_2$, where the subscripts denote the $R$ charges and
$\mathcal{M}$ is an arbitrary scalar function of \SU5 invariant mass
parameters $m$ and \SU5 breaking VEVs $\langle H_0 \rangle$ (and a `cut--off'
scale $\Lambda$).

In what follows, we will show that it is impossible to:
\begin{enumerate}
\item spontaneously break $\SU5\to G_\mathrm{SM}$ by assigning a VEV to a
suitable representation,
\item keep the $R$ symmetry unbroken and to
\item avoid extra massless $G_\mathrm{SM}$ charged representations 
\end{enumerate}
at the same time. We will present our analysis in two steps. First, we focus on
the simplest possibility of spontaneously breaking $\SU5 \times \Z{M}^R \to
G_\mathrm{SM} \times \Z{M}^R$ by giving a VEV to a $\rep{24}$--plet and allowing
only for further $\rep{24}$--plets as mass partners in the model. This setting
already illustrates the crucial point of our proof, namely the obstruction to
decouple unwanted exotics. In the second step, we will discuss the general case
where the symmetry is broken by an arbitrary reducible representation and where
we allow for arbitrary further representations to render all exotics massive.

\subsubsection{Breaking the GUT symmetry using only $\rep{24}$--plets}
\label{sec:Step1}

Since we wish to leave $\Z{M}^R$ unbroken, the $\rep{24}$--plet that is
supposed to break the GUT symmetry to the SM group has to carry $\Z{M}^R$
charge 0. The branching rule
for $\rep{24}$ is
\begin{equation}
 \rep{24}~=~(\rep{8},\rep{1})_0\oplus
 (\rep{1},\rep{3})_0\oplus
 (\rep{1},\rep{1})_0\oplus
 (\rep{3},\rep{2})_{\frac{-5}{6}}\oplus
 (\crep{3},\rep{2})_{\frac{5}{6}}\;.
\end{equation}
In the course of spontaneous symmetry breakdown the last two SM representations 
$(\rep{3},\rep{2})_{\frac{-5}{6}}\oplus(\crep{3},\rep{2})_{\frac{5}{6}}$ get
absorbed in the longitudinal components of the extra gauge bosons. However, we
are now left with chiral superfields transforming as $(\rep{8},\rep{1})_0$ and $
(\rep{1},\rep{3})_0$ and carrying $R$ charge 0. The crucial observation here is
that the mass term  $m\,\rep{24}\,\rep{24}$ for the adjoint is forbidden:
although the $\rep{24}$--plet is a real \SU5 representation the mass term is
prohibited by the $\Z{M}^R$ symmetry because $0+0 \not\equiv 2\pmod{M}$.
Therefore, in order to give masses to the extra $(\rep{8},\rep{1})_0$ and
$(\rep{1},\rep{3})_0$ fields, we would have to introduce further fields
furnishing the same representations and carrying $R$ charge 2. Yet we cannot
simply introduce these desired SM representations, rather we have to add 
complete \SU5 multiplets. That is, we have to introduce one or more multiplets
that contain $(\rep{8},\rep{1})_0$ and $ (\rep{1},\rep{3})_0$ and carry $R$
charge 2. Here, in the first step, we consider the possibility to add a
$\rep{24}$--plet with $R$ charge 2, the $\rep{24}$ being the smallest multiplet
containing $(\rep{8},\rep{1})_0$ and/or $(\rep{1},\rep{3})_0$. While this, in
principle, allows us to write mass terms for $(\rep{8},\rep{1})_0$ and $
(\rep{1},\rep{3})_0$, we are now left with extra chiral fields transforming as
$(\rep{3},\rep{2})_{-\frac{5}{6}}$ and $(\crep{3},\rep{2})_{\frac{5}{6}}$ and
carrying $R$ charge 2. Now of course, we may add another $\rep{24}$--plet with
$R$ charge 0, but this will lead us just back to the problem we started with:
extra massless $(\rep{8},\rep{1})_0$ and $ (\rep{1},\rep{3})_0$ representations.
So we conclude that  adding an arbitrary but finite number of
$\boldsymbol{24}$--plets with $R$ charges $0$ or $2$ cannot solve the problem;
we will always obtain massless exotics.

\subsubsection{General case}
\label{sec:Step3}

Could one rectify the situation by introducing representations different
from $\rep{24}$ as the GUT breaking Higgs and as mass partners? In what follows,
we will show that this is not the case.

Instead of using just a $\rep{24}$ to break \SU5 to the SM,
we will use an arbitrary, finite,
possibly reducible representation $\rep{H}_0$, such as one or
several
$\rep{75}$--plets. This representation $\rep{H}_0$ has to fulfill two
requirements: 
\renewcommand{\labelenumi}{(\roman{enumi})}
\begin{enumerate}
  \item it has $\Z{M}^R$ charge 0 (as suggested by the subscript) in
order to leave this symmetry unbroken and
  \item it lies within the congruence class
\cite{Lemire:1980,Slansky:1981yr} of the $\rep{24}$.
\end{enumerate}
The second property must hold because only SM singlet components of \SU5
representations may attain VEVs. They can only originate from the adjoint
congruence class because of the following reasoning. The decomposition of an
$\SU5$ representation into SM representations can be accomplished by
using an invertible projection matrix $P$ \cite[section~6]{Slansky:1981yr},
\begin{equation}
  P=\begin{pmatrix}
 1 & 1 & 0 & 0 \\
 0 & 0 & 1 & 1 \\
 0 & 1 & 1 & 0 \\
 -\frac{1}{3} & \frac{1}{6} & -\frac{1}{6} & \frac{1}{3}
    \end{pmatrix},
\end{equation}
which maps \SU5 weights (in the Dynkin basis) to the corresponding SM
weights. \SU5 has five
congruence classes of mutually disjoint weight lattices. Using the projection
$P$, one can map the weight lattice of each congruence class onto 
an equivalence class of SM weights. (Of course, there are further SM
representations that do not fit into complete \SU5 multiplets.) 
These equivalence classes are disjoint.
The SM singlet lies in the class that originates from the \SU5
congruence class of the adjoint, and also the representation
$(\rep{3},\rep{2})_{\frac{-5}{6}}\oplus (\crep{3},\rep{2})_{\frac{5}{6}}$, which
is needed to make the extra \SU5 gauge bosons massive, lies in the same class.
Hence we can choose $\rep{H}_0$ from the adjoint \SU5 congruence class. Of
course, there may be additional fields with $R$ charge 0 that do not obtain a
VEV and therefore do not have to be in this class, but they do not interfere
with the following arguments.

In order to arrive at the precise SM spectrum, we allow for an
additional finite, possibly reducible representation $\rep{R}_2$ with $R$ charge
2. All non--trivial SM representations contained in $\rep{H}_0$ except for
$(\rep{3},\rep{2})_{\frac{-5}{6}}\oplus
(\crep{3},\rep{2})_{\frac{5}{6}}$ have to obtain masses by pairing
up with representations in $\rep{R}_2$ in order to avoid massless exotics.
However, in the following we will show that this is impossible.

Without loss of generality, we can restrict $\rep{R}_2$ to belong to the
\SU5 congruence class of the adjoint. This is because the complex conjugates of
representations from the adjoint congruence class lie in the same class.
Therefore no SM representation coming from a different congruence class can pair
up with representations coming from $\rep{H}_0$. Clearly, one can also remove
those representations from $\rep{H}_0$ and $\rep{R}_2$ for which one can write
down $\SU5 \times \Z{M}^R$ invariant mass terms. For notational simplicity we
will call the remaining representations again $\rep{H}_0$ and $\rep{R}_2$,
respectively.

Now we take the highest weights $\Lambda_0$ from $\rep{H}_0$
and $\Lambda_2$ from $\crep{R}_2$,
respectively.\footnote{The term highest weight can be defined using the
following ordering: $\lambda>\mu$ if and only if the first non--zero coefficient
$n_i$ in the expansion $\lambda-\mu = \sum_i{n_i\alpha_i}$, where $n_i \in
\mathbb{N}_0$ and
$\alpha_i$ are the simple roots, is greater than zero.} They cannot be equal
because otherwise the corresponding representations could pair up and,
therefore, would have been removed in the previous step. Thus we arrive at
two cases:
(i) $\Lambda_0>\Lambda_2$ and (ii) $\Lambda_2>\Lambda_0$.

\paragraph{Case 1: \boldmath$\Lambda_0>\Lambda_2$\unboldmath.}  Using the
projection matrix $P$, the \SU5 representation with highest weight
$\Lambda_0=(a_1,a_2,a_3,a_4)$ introduces an SM representation
$\rep{r}=\rep{r}(P(\Lambda_0))$ with highest weight $P(\Lambda_0)=P \cdot
(a_1,a_2,a_3,a_4)^\mathrm{T}$. In what follows, we will show that (i) $\rep{r}$
is neither any of the desired representations $(\rep{3},\rep{2})_{\frac{-5}{6}}$
or $(\crep{3},\rep{2})_{\frac{5}{6}}$ nor (ii) MSSM matter nor (iii) an SM
singlet and that (iv) it cannot pair up with any partner from $\rep{R}_2$.
\renewcommand{\labelenumi}{(\roman{enumi})}
\begin{enumerate}
\item Since $P$ establishes a one--to--one correspondence between
\SU5 and SM weights, we can use its inverse to calculate the inverse image of
the highest weight of
$(\rep{3},\rep{2})_{\frac{-5}{6}}=((1,0),(1),\frac{-5}{6})$,
\begin{equation}
  P^{-1}\cdot (1,0,1,\frac{-5}{6})^\mathrm{T}~=~(1,0,1,-1)^\mathrm{T}\;,
\end{equation}
which is not a highest weight of \SU5. Hence it cannot be equal to $\Lambda_0$.
The same holds for the highest weight
of $(\crep{3},\rep{2})_{\frac{5}{6}}$. Therefore $\rep{r} \neq
(\rep{3},\rep{2})_{\frac{-5}{6}}$ and $\rep{r} \neq
(\crep{3},\rep{2})_{\frac{5}{6}}$. 

\item $\rep{r}$ can also not be part of the
SM matter content because $\rep{H}_0$ was chosen to be in the congruence class
of the adjoint and matter originates from different classes. 

\item Furthermore,
we can exclude the case that $\rep{r}$ is an SM singlet because otherwise the
highest weight would be $\Lambda_0=(0,0,0,0)$ and we could not give masses to
the extra \SU5 gauge bosons. 

\item In addition to that, there is no chance that
$\rep{r}$ can pair up because the necessary partner $\crep{r}$ is not contained
in $\rep{R}_2$. This is true because, by assumption, the highest weight
$\Lambda_2$ of $\crep{R}_2$ is smaller than $\Lambda_0$.
\end{enumerate}
Therefore $\rep{r}$ is a massless, SM charged exotic.

\paragraph{Case 2: $\boldsymbol{\Lambda_2>\Lambda_0}$.}
For analogous reasons as in case 1, $\rep{r}' =
\overline{\rep{r}(P(\Lambda_2))}$ can neither be SM matter nor a singlet nor can
it pair up with a partner from $\rep{H}_0$ to obtain a mass. As it originates
from $\rep{R}_2$, its $R$ charge is 2 and therefore it can also not be used to
give masses to the extra \SU5 gauge bosons. Again we are left with at least one
massless, SM charged exotic in representation $\rep{r}'$.

\parskip 12pt
Altogether we have seen that, if one wants to break \SU5 to the SM with a
finite number of multiplets while leaving a $\Z{M}^R$ unbroken, one will
necessarily obtain massless, SM charged exotics.
\parskip 0pt

Let us illustrate the main point with an easy example, based on
$\rep{H}_0=\rep{24}$ and $\rep{R}_2=\rep{75}$ (such that
$\crep{R}_2=\rep{R}_2$). The highest weights of the two sets are
$\Lambda_0=(1,0,0,1)$ and $\Lambda_2=(0,1,1,0)$, respectively. Out of the two
highest weights, $\Lambda_2$ is the higher one and we are left with a massless,
SM charged field $\rep{r}(P(0,1,1,0))= (\rep{8},\rep{3})_0$ with $R$ charge 2.
There are, of course, further massless exotic states.

At this point, a remark is in order. The restriction to a finite number of
multiplets is crucial for our proof. Our analysis is, in this sense, very
similar to the one by Goodman and Witten~\cite{Goodman:1985bw}, where
obstructions for building 4D GUT models with a finite number of multiplets have
been identified.  As discussed in \cite{Goodman:1985bw} and as we shall see
explicitly in section~\ref{sec:ExtraDims}, in theories with compact extra
dimensions, which from a 4D perspective appear to have infinitely many states,
our no--go theorem does not apply.

\subsection{Further no--go theorems}
\label{sec:furtherNoGo}

It is straightforward to extend the no--go theorem to the case of singlet
extensions of the MSSM as well as to GUTs with gauge groups containing \SU5 as a
subgroup (such as \SO{10}).

\subsubsection{No--go for singlet extensions of the MSSM}

As already mentioned in the introduction, our arguments also
apply to the case of singlet extensions of the MSSM. This is because the
presence of additional singlets cannot lead to a decoupling of the charged
states. Therefore we will still be left with charged light states
beyond the MSSM spectrum.

\subsubsection{No--go for GUTs with simple gauge group 
$\boldsymbol{G\supset\SU5}$}
\label{sec:LargerGUTs}

In the case of a GUT with simple gauge group $G$ containing \SU5 as a subgroup,
the multiplets will become larger and the constraints derived in
section~\ref{sec:NoGoZM} get tighter. To see this, one can decompose all
representations of $G$ into irreducible representations with respect to the \SU5
subgroup. Adding representations of $G$ can therefore not circumvent our 
no--go theorem. One may now wonder whether the extra gauge bosons from $G/\SU5$ 
may provide mass partners for the unwanted exotics discussed in 
section~\ref{sec:Step3}. However, these gauge bosons come in \SU5 congruence 
classes which are different from the one containing the adjoint. This is 
because the difference between weights of extra gauge bosons and a weight of a 
representation in the \SU5 adjoint congruence class is not an \SU5 root. (For 
instance, in $\SO{10}/\SU5$ one has extra gauge bosons transforming as
$\rep{10}\oplus\crep{10}$ while in the case of \SU6 one gets extra 
$\rep{5}\oplus\crep{5}$ states.) Therefore the extra gauge bosons 
from $G/\SU5$ cannot pair up with the unwanted exotics discussed in 
section~\ref{sec:Step3} and thus cannot interfere with our proof. Hence our 
no--go theorem from section~\ref{sec:NoGoZM} applies to the case of a GUT 
based on a simple group $G\supset\SU5$ as well. In particular, it also holds 
in the case of $G=\SO{10} \supset \SU5\times \U1_{\chi}$ and therefore 
excludes $R$ symmetries in another important class of 4D GUT models.

\section{Implications for model building}
\label{sec:applications}

As already stated in the introduction, assuming that (in the effective MSSM
theory) matter is contained in GUT multiplets and that the theory is
anomaly--free, only $R$ symmetries can control the $\mu$ term
\cite[section~2.1]{Lee:2011dya}. 
(The role of anomaly--free discrete $R$ symmetries in controlling the $\mu$
parameter has also been discussed earlier in \cite{Babu:2002tx}).
 Yet, as we have shown, such symmetries are not
available if the MSSM is to be completed by a 4D GUT. Therefore it is not
possible to obtain a `natural' (in 't Hooft's sense), i.e.\ symmetry--based,
solution to the doublet--triplet splitting problem in four dimensions.

This applies in particular to the five $\Z{M}^R$ symmetries recently discussed
in \cite{Lee:2011dya}. These are the only family--independent, anomaly--free
symmetries for the MSSM which (i) commute with \SU5 in the matter sector, (ii)
forbid the $\mu$ term at tree level, (iii) allow for the usual Yukawa couplings
and the dimension five neutrino mass operator and (iv) suppress proton decay. 
Our no--go theorems tell us that these symmetries, providing simple and
simultaneous solutions to the $\mu$ and proton decay problems, are not available
in 4D GUT model building.

\section{$\boldsymbol{\Z4^R}$ MSSM from GUTs in extra dimensions}
\label{sec:ExtraDims}

As mentioned above, our no--go theorems do not apply in the presence of extra
dimensions, where new ways of GUT symmetry breaking arise
\cite{Witten:1985xc,Breit:1985ud}. Let us discuss the case of breaking by a
discrete Wilson line. This Wilson line breaks the GUT symmetry in the same way
as an adjoint VEV would do, i.e.\ $\SU5\to G_\mathrm{SM}$ in the
phenomenologically interesting case. However, a $\Z2$ (or more generally a
$\Z{N}$) Wilson line is quantized. Hence there are no continuous deformations
(i.e.\ $(\rep{8},\rep{1})_0$ or $(\rep{1},\rep{3})_0$ fields). From the 4D point
of view, the symmetry breaking is not spontaneous. Or, adopting the point of
view suggested in \cite{Goodman:1985bw}, there are infinitely many
representations such that each of the unwanted states can find a mass
partner to pair up. Therefore this mechanism evades our no--go theorems.

Wilson line breaking of the GUT symmetry has been implemented  in the context of
MSSM Calabi--Yau compactifications  \cite{Braun:2005ux,Bouchard:2005ag}. More
recently, it has also been realized in heterotic orbifold compactifications
\cite{Blaszczyk:2009in}. At this point it is worthwhile to point out that there
is a slightly confusing terminology. What is traditionally called a ``discrete
Wilson line on an orbifold'' \cite{Ibanez:1986tp} is in fact a discrete Wilson
line on the underlying torus and a difference between ``local shifts'' on the
orbifold (see \cite{GrootNibbelink:2003rc} for an explanation of local shifts). 
An appealing feature of the orbifold models  is that there the discrete $R$
symmetries are not imposed by hand, rather they originate from the Lorentz
symmetry of compact dimensions~\cite{Kappl:2010yu}, and their appearance can be
related to the fact that orbifolds are highly symmetric compactifications.
More importantly for phenomenology, it has been demonstrated explicitly
that the remnant $\Z{M}^R$ symmetries can be of the type discussed above.
Specifically, in a global $\Z2\times\Z2$ orbifold model vacua with the precise
MSSM spectrum and a residual $\Z4^R$ symmetry have been identified. This $\Z4^R$
is the unique $\Z{M}^R$ symmetry for which  the discrete matter charges commute
with \SO{10} \cite{Lee:2010gv}. It forbids the $\mu$ term and
dimension five proton decay operators at tree level, and contains matter parity
as a subgroup. In summary, we see that grand unified theories in extra
dimensions allow us to circumvent
our no--go theorems (and, arguably, provide us with the most compelling way
of doublet--triplet splitting). In the context of heterotic orbifolds it is
rather straightforward to realize the phenomenologically attractive $\Z4^R$ in
MSSM vacua, and, moreover one obtains a simple geometric intuition for
how this discrete $R$ symmetry emerges.

\section{Summary}
\label{sec:summary}

We have shown that 4D GUTs cannot provide an ultraviolet completion of the
MSSM with a residual $\Z{M\geq3}^R$ symmetry, nor with a continuous $R$
symmetry. These theories fail because, as we demonstrated, one will necessarily
have additional SM charged states at low energies.

Given that, assuming (i) matter charges that commute with \SU5 and (ii) anomaly
freedom, only $R$ symmetries can forbid the $\mu$ term in the MSSM, we have
argued that it is not possible to obtain a `natural', i.e.\ symmetry--based,
solution to the doublet--triplet problem in four dimensions. In particular, none
of the five generation--independent, anomaly--free discrete $R$ symmetries,
which forbid the $\mu$ term and suppress proton decay in the MSSM, can be
implemented in a 4D GUT (based on a simple gauge group).

On the other hand, as we have discussed, higher--dimensional models of
grand unification (with an explicit string completion) can give us
precisely the MSSM with a residual $\Z4^R$ symmetry. In such models, the
doublet--triplet splitting has a very simple solution and the $\mu$
parameter is related to the gravitino mass. In these constructions, the
discrete $R$ symmetries are not imposed by hand, rather they originate from the
Lorentz symmetry of compact dimensions.

\subsection*{Acknowledgments}

We would like to thank Rolf Kappl and Stuart Raby for useful discussions. This
research was supported by the DFG cluster of excellence Origin and Structure of
the Universe by Deutsche Forschungsgemeinschaft (DFG). This material is partly
based upon work done at the Aspen Center for Physics and supported by the
National Science Foundation under Grant No.\ 1066293. One of us (M.R.) would
also like to thank the CERN theory group for hospitality.

\bibliography{Orbifold}
\bibliographystyle{NewArXiv}

\end{document}